\documentclass[sigconf]{acmart}
\AtBeginDocument{%
  \providecommand\BibTeX{{%
    \normalfont B\kern-0.5em{\scshape i\kern-0.25em b}\kern-0.8em\TeX}}}

\usepackage{graphicx}
\usepackage{amsmath}
\usepackage{makecell}
\usepackage{multirow}
\usepackage{amsthm}
\usepackage{booktabs}
\usepackage{algorithm}
\usepackage{algorithmic}
\usepackage{placeins}
\usepackage{xcolor}
\usepackage{balance}

\copyrightyear{2021}
\acmYear{2021}
\setcopyright{acmlicensed}\acmConference[CIKM '21]{Proceedings of the 30th ACM International Conference on Information and Knowledge Management}{November 1--5, 2021}{Virtual Event, QLD, Australia}
\acmBooktitle{Proceedings of the 30th ACM International Conference on Information and Knowledge Management (CIKM '21), November 1--5, 2021, Virtual Event, QLD, Australia}
\acmPrice{15.00}
\acmDOI{10.1145/3459637.3482095}
\acmISBN{978-1-4503-8446-9/21/11}

\begin{document}
\fancyhead{}

\title{Embedding Node Structural Role Identity Using Stress Majorization}

\author{Lili Wang}
\affiliation{%
  \institution{Dartmouth College}
  \city{Hanover}
  \state{New Hampshire}
  \country{USA}}
\email{lili.wang.gr@dartmouth.edu}

\author{Chenghan Huang}
\affiliation{%
  \institution{Millennium Management, LLC}
  \city{New York}
  \state{New York}
  \country{USA}}
\email{njhuangchenghan@gmail.com}

\author{Weicheng Ma}
\affiliation{%
  \institution{Dartmouth College}
  \city{Hanover}
  \state{New Hampshire}
  \country{USA}}
\email{weicheng.ma.gr@dartmouth.edu}

\author{Ying Lu}
\affiliation{%
  \institution{Stony Brook Univeristy		}
  \city{Stony Brook}
  \state{New York}
  \country{USA}}
\email{luying830@gmail.com}

\author{Soroush Vosoughi}
\affiliation{%
  \institution{Dartmouth College}
  \city{Hanover}
  \state{New Hampshire}
  \country{USA}}
\email{soroush.vosoughi@dartmouth.edu}
\linespread{0.91} 
\begin{abstract}

Nodes in networks may have one or more functions that determine their role in the system. As opposed to local proximity, which captures the local context of nodes, the role identity captures the functional ``role" that nodes play in a network, such as being the center of a group, or the bridge between two groups. This means that nodes far apart in a network can have similar structural role identities. Several recent works have explored methods for embedding the roles of nodes in networks. However, these methods all rely on either approximating or indirect modeling of structural equivalence. In this paper, we present a novel and flexible framework using stress majorization, to transform the high-dimensional role identities in networks \emph{directly} (without approximation or indirect modeling) to a low-dimensional embedding space. Our method is also flexible, in that it does not rely on specific structural similarity definitions. We evaluated our method on the tasks of node classification, clustering, and visualization, using three real-world and five synthetic networks. Our experiments show that our framework achieves superior results than existing methods in learning node role representations.

\end{abstract}





\maketitle

\section{Introduction}
Network (or graph) embedding, which involves learning low dimensional feature representations of nodes and links, has in recent years become a popular topic of research. Among them, structural role embedding is one type of embedding method that focuses on identifying nodes serving different ``functions" in a network (e.g., acting as a bridge between two communities or being the center of a community). Different from local proximity (the focus of methods like DeepWalk \cite{deepwalk} and node2vec \cite{node2vec}), nodes far apart in a network and having different local contexts can be similar in their structural role identity. Several structural role embedding methods have been proposed in recent years. Among them, struc2vec \cite{struc2vec} and GraphWave \cite{graphwave} are two representative methods. 

However, current methods approach node structural role embedding either through indirect modeling (such as GraphWave) or non-precise methods (such as struc2vec). For example, struc2vec constructs a weighted multi-layer graph that uses random walks to generate a context for each node, which is then fed into a language model. Basically, their model is actually based on the assumption that two nodes are structurally similar if and only if the context generated by the random walk is similar. However, the randomness of random walk makes the embeddings imprecise, often leading to nodes with the exact same role having different (though admittedly, similar) embeddings. 

GraphWave, on the other hand, defines a wavelet coefficient matrix $\Psi$ and uses the distribution of energy that comes from other nodes to model node roles. Though this method can perfectly preserve the roles of nodes, it cannot capture subtle dissimilarities between roles, given that is based on indirect modeling of node roles (i.e., the energy distribution of nodes). Furthermore, this indirect modeling approach is not flexible as it relies on specific structural similarity definitions based on the energy distributions of the nodes.

We propose a direct and precise (without approximation or indirect modeling) embedding method for node structural role identity using stress majorization. Though perfectly preserving role similarities between nodes in embedding space is impossible (since embedding reduces the dimensions, leading to inevitable information loss), our method minimizes this information loss. Moreover, our method is also flexible, in that it does not rely on specific structural similarity definitions. Specifically, in this paper, we make the following contributions: 
\begin{itemize}

\item We present a novel and flexible structural embedding framework, using stress majorization, that can directly and precisely capture the role structural identities and similarities of nodes in networks. Our method is also flexible, in that it does not rely on specific structural similarity definitions.

\item We prove mathematically that our method embeds nodes with the same roles into the exact same position in the embedding space.

\item We evaluate our method on the fundamental tasks of node classification, clustering, and visualizations on three real-world and five synthetic networks. Our experiments show that our framework outperforms existing methods in learning node role representations.

\end{itemize}


\section{Related Work}
As mentioned in the introduction, microscopic-structure preserving embedding methods \cite{deepwalk,node2vec,line,wang2021embedding,wang2021hyperbolic} can not capture the roles of nodes in networks.
Besides struc2vec and GraphWave, discussed in the introduction, struc2gauss \cite{struc2gauss} is a newer method for structural role preserving embedding. struc2gauss first generates structural context for each node and then learns representations from Gaussian embeddings. A Gaussian distribution is used to represent each node: the mean is used to represent the position and the covariance is used to represent the uncertainty. This method, like struc2vec, is also imprecise. There are also several related methods that focus on concepts related to node structural role embedding. DRNE \cite{DRNE} introduces a concept similar to structural roles called regular equivalence and uses a layer normalized LSTM \cite{hochreiter1997long} to learn the representations of nodes through aggregating their neighborhoods in a recursive way. RolX \cite{rolx} gives a mixed-membership approach that uses non-negative matrix-factorization to assign every node a distribution over the set of identified roles. Our own prior work tackles embedding node role identity into hyperbolic space \cite{wang2020embedding} and embedding role identity over time \cite{wang_tois_2021}. Finally, SNS \cite{SNS} uses graphlets for structural similarity, and combines neighborhood information and local subgraphs similarity to learn embeddings.

\section{Framework}
Let $G=(V, E)$ be an undirected and unweighted network, where $V$ is a set of vertices, and $E \subseteq V \times V $ is the set of unweighted edges between vertices in $V$. We consider the problem of representing all the nodes in $V$ as a set of d-dimensional vectors $X_i \in \mathbb{R}^d$, with $d<<|V|.$ Our framework consists of two parts: calculation of the structural role distances of nodes and the use of stress majorization to generate embeddings.
\subsection{Calculation of Structural Role Distance}
The structural role distance in our model can use any of the known measurements. In this paper, we use the similarity of $k-$hop degree sequence defined by struc2vec. Let $L_{k}(u)$ denote the ordered degree sequence of the nodes at exactly $k$ hop count from $u$ in $G$.
The structural role similarity of two nodes $u$ and $v$ considering their $k$ hop neighbors can be defined as the similarity of the two ordered sequences $L_{k}(u)$ and $L_{k}(v)$. Since these two sequences may not have equal sizes, we use \emph{Fast Dynamic Time Warping} (FastDTW) \cite{fastdtw} to
measure the distance between two ordered degree sequences. FastDTW is able to find the optimal alignment between two arbitrary length time series and limits both the time and space complexity to $O(n)$. Since elements of the sequences $L_{k}(u)$ and $L_{k}(v)$ are degrees of nodes, we adopt the following 
cost function of $i$th and $j$th elements in the above two sequences for FastDTW:
\begin{equation}
\small
    cost(L_{k}^i(u), L_{k}^j(v)) =\frac{\max (L_{k}^i(u), L_{k}^j(v))}{\min (L_{k}^i(u), L_{k}^j(v))}-1 
    \label{distance}
\end{equation}
Instead of measuring the absolute difference of degrees, this distance metric measures the relative difference which is more suitable for degree differences (since the degrees can be arbitrarily large). 
The structural distance of two nodes $u$ and $v$ considering their $k$-hop neighborhoods can be defined as:
\begin{equation}
\small
       dis_{k}(u,v)=\sum_{i=0}^k{ w_{k}\cdot FastDTW(L_{i}(u),L_{i}(v))}
\end{equation}

Where $w_{k}$ is the importance weight of each hop. In our experiment, we set all the $w_{k}$ to be equal. When we set the $k$  to be the diameter of the graph, $dis_{k}(u,v)$ depicts the structural role distance of these two nodes.

\subsection{Stress Majorization}
After we calculate the pairwise structural role distances, our goal is to embed the nodes into low dimensional space and make the pairwise distances of the nodes equal or close to the structural role distances. Notice that the structural role distances may not obey triangle inequality, which means that classical Multidimensional Scaling (MDS) \cite{torgerson1958theory} cannot be used and thus we adopt stress majorization \cite{kruskal1964multidimensional,borg2005modern} here. We use an $n \times d$ matrix $X$ to represent the $d$-dimensional embedding vectors, with the row vectors $X_i \in \mathbb{R}^d$ . The $n \times n$ matrix $D$ is the structural role distance matrix, with $D_{i,j}=dis_{k}(i,j)$. Given these definitions, we define the stress function as
\begin{equation}
\small
    stress(X) = \sum_{i<j}{\left(\left\Vert X_i-X_j\right\Vert-D_{i,j}\right)^2}.
\end{equation}
The problem can be formulated as seeking a matrix $X$ to minimize the stress function given $D_{i,j}$. The theorem and corresponding proof below give a bound of the stress function.
\begin{theorem}
The Laplacian matrix $L$ is defined as
\small
$$
L_{i,j}=\left\{
\begin{aligned}
& -1 \,\,\, & i \neq j \\
& n-1 \,\,\, & i=j  \\
\end{aligned}\,\,\,.
\right.
$$
For any $n \times d$ matrix $Y$, we define matrix $L^Y$ as
$$
\small
L^Y_{i,j}=\left\{
\begin{aligned}
& -\frac {D_{i,j}}{\left\Vert Y_i-Y_j\right\Vert} \,\,\, & i \neq j \\
& -\sum_{k \neq i} {L^Y_{i,k}} \,\,\, & i=j  \\
\end{aligned}\,\,\,,
\right.
$$
where $Y_i$ is the $i$-th row vector of matrix $Y$. And the function $f^Y(X)$ is defined as
\begin{equation}
\small
    \sum_{i<j}{D_{i,j}^2} + tr(X^TLX) - 2tr({X^T}{L^Y}Y).
\end{equation}
We must have
\begin{equation}
\small
    f^Y(X) \geq stress(X).
\end{equation}
The equality holds when $X=Y$.
\end{theorem}
\begin{proof}
Expanding the definition of stress function and we can get $stress(X)$ as
\begin{equation}
\small
     \sum_{i<j}{D_{i,j}^2} + \sum_{i<j}{{\left\Vert X_i-X_j\right\Vert}^2} - 2\sum_{i<j}{D_{i,j}{\left\Vert X_i-X_j\right\Vert}}.
\end{equation}
Notice that the second term is a quadratic form, which can be written in matrix form:
\begin{equation}
\small
     \sum_{i<j}{{\left\Vert X_i-X_j\right\Vert}^2}=tr(X^TLX).
\end{equation}
In this way, we just have to show that
\begin{equation}
\small
    \sum_{i<j}{D_{i,j}{\left\Vert X_i-X_j\right\Vert}} \geq tr({X^T}{L^Y}Y).
\end{equation}
According to Cauchy-Schwartz inequality,
\begin{equation}
\small
    {\left\Vert X_i-X_j\right\Vert}{\left\Vert Y_i-Y_j\right\Vert} \geq (X_i-X_j)^T(Y_i-Y_j).
\end{equation}
Therefore, the third term can be bounded as follows
\begin{equation}
\small
    \sum_{i<j}{D_{i,j}{\left\Vert X_i-X_j\right\Vert}} \geq \sum_{i<j}{\frac {D_{i,j}(X_i-X_j)^T(Y_i-Y_j)}{\left\Vert Y_i-Y_j\right\Vert}}.
\end{equation}
Write the right-hand side in matrix form, we get
\begin{equation}
\small
    \sum_{i<j}{\frac {D_{i,j}(X_i-X_j)^T(Y_i-Y_j)}{\left\Vert Y_i-Y_j\right\Vert}}=tr({X^T}{L^Y}Y).
\end{equation}
Combining (10) and (11), we get (8), and thus the theorem has been proved.
\end{proof}

For a given $Y$, $f^Y(X)$ is minimized when $\frac {\partial f^Y(X)}{\partial X}=0$. This means that we need to solve 
\begin{equation}
\small
    LX={L^Y}Y.
\end{equation}
For a given layout $X(t)$, we take $X(t+1)$ which makes
\begin{equation}
\small
    LX(t+1)={L^{X(t)}}X(t).
\end{equation}
If $X(t+1) \neq X(t)$, we must have
\begin{equation}
\small
stress(X(t))=f^{X(t)}(X(t))
>f^{X(t)}(X(t+1))
\geq stress(X(t+1))
\end{equation}
Now we can design an iterative optimization process as shown below:\\
$\textbf{Step\,1}$ \,\,Initiate $X(0)$.\\
$\textbf{Step\,2}$ \,\,Calculate $X(t+1)$ from $X(t)$ by solving $LX(t+1)={L^{X(t)}}X(t)$.\\
$\textbf{Step\,3}$ \,\,Check whether the error is below tolerance. If
$$
\left| \frac {stress(X(t+1))-stress(X(t))}{stress(X(t))}\right|<\epsilon,
$$
terminate the process, else go back to $\textbf{Step\,2}$. Typically, we set $\epsilon=10^{-3}$. Run this algorithm until it converges. The final $X$ is the embedding matrix that we want.

\subsection{Proof of Structurally Equivalent Nodes}
In this section, we prove the claim we proposed in the introduction, which is that nodes with the same roles will be embedded into the exact same position. 
\begin{theorem}
Assume that we have $i_0$ and $j_0$, with $D_{i_0,k}=D_{j_0,k}$ for any $k \neq i,j$, and $D_{i_0,j_0}=0$. The result of stress majorization must follow $X_{i_0} = X_{j_0}$.
\end{theorem}
\begin{proof}
If $X_{i_0} \neq X_{j_0}$, without loss of generality, we can assume $i_0<j_0$ and
\begin{equation}
\small
\sum_{k \neq i_0,j_0}{\left(\left\Vert X_{i_0}-X_k\right\Vert-D_{i_0,k}\right)^2} \\\leq \sum_{k \neq i_0,j_0}{\left(\left\Vert X_{j_0}-X_k\right\Vert-D_{j_0,k}\right)^2}.
\end{equation}
Let us construct a new solution $\widetilde X$, with row vectors following
\begin{equation}
\small
\widetilde X_k=\left\{
\begin{aligned}
& X_k \,\,\, & k \neq i_0\,\,& and\,\,k \neq j_0 \\
& X_{i_0} \,\,\, & k = i_0\,\,& or\,\,k = j_0 \\
\end{aligned}\,\,\,.
\right.
\end{equation}
It is obvious that
\begin{equation}
\small
 \sum_{\begin{subarray}{c}(i,j) \neq (i_0,j_0)\\i<j\end{subarray}}{\left(\left\Vert X_i-X_j\right\Vert-D_{i,j}\right)^2} \geq
 \sum_{\begin{subarray}{c}(i,j) \neq (i_0,j_0)\\i<j\end{subarray}}{\left(\left\Vert \widetilde X_i-\widetilde X_j\right\Vert-D_{i,j}\right)^2}.
\end{equation}
Considering $\left\Vert X_{i_0}-X_{j_0}\right\Vert>0=\left\Vert \widetilde X_{i_0}-\widetilde X_{j_0}\right\Vert$, we get
$$
stress(X(t))>stress(\widetilde X(t)),
$$
which contradicts with the definition of stress majorization. Therefore, the same roles must be embedded into the same position.
\end{proof}

\section{Experiment}
In this section, we evaluate and compare our method on qualitative (visualization) and quantitative (node classification and clustering) tasks on real-world and synthetic networks. 

\begin{figure}[ht]
\centering
\includegraphics[width=0.95\linewidth]{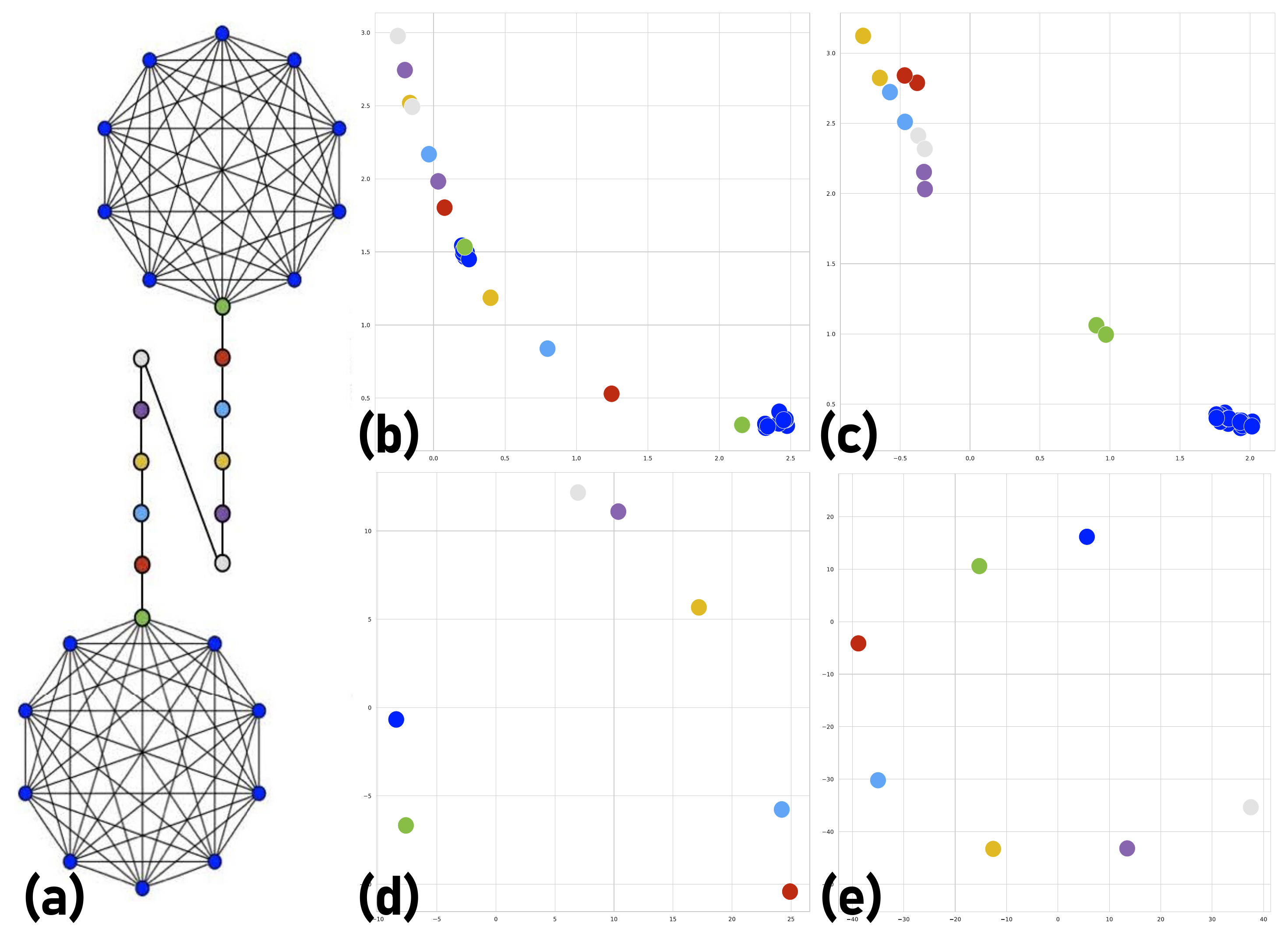}
\centering

\caption{(a) The Barbell graph (from \cite{struc2vec}) used in our experiment. Visualization of embeddings of nodes in the barbell graph by (b) node2vec (c) struc2vec (d) GraphWave (e) ours. }
\label{barbell}
\end{figure}
\subsection{Node Structural Role Embedding Visualization}
The first set of experiments involves qualitative evaluation of the generated embeddings through visualization on a synthetic Barbell graph. The barbell graph consists of two complete subgraphs connected by a ``bridge''. Figure \ref{barbell}(a) shows the barbell graph used in our experiments, where each subgraph has 10 nodes and the bridge has a length of 11. In the figure, the structurally equivalent nodes
have the same color. Figure \ref{barbell}(b) shows the embeddings generated by node2vec, which is a representative example of local proximity methods. As can be seen, node2vec does not capture structural role identities. This is to be expected as node2vec generates embeddings based on the local context of nodes (this can be seen in the figure, where the nodes of the two complete subgraphs are placed separately and next to the nodes in the bridge that are close to them). Figures \ref{barbell}(c), (d), and (e), show the results of the three structural role embedding methods, struc2vec, GraphWave, and our method, respectively. We use struc2vec as a representative method of all the non-precise structural role embedding methods (like struc2gauss and DRNE). Compared to our method and GraphWave, struc2vec cannot embed structural identical nodes precisely. This is because, as explained earlier, struc2vec uses the context generated by random walk to model similarity, which makes the embeddings stochastic and imprecise. GraphWave and our method both can embed the nodes with the same role precisely (as shown by the overlap of nodes with the same color). However, whereas our model perfectly captures the distance between nodes in the ``bridge'' (grey, purple, yellow, teal, red, and green nodes), GraphWave does not. For instance, the grey-purple and teal-red pairs are embedded much closer to each other than the purple-yellow and yellow-teal pairs, which is not correct. The pairs are correctly positioned with similar distances to each other by our method. Also, the green nodes have a special role as they are the connections between the subgraphs and the bridge. Our method correctly captures all the role information of green nodes. First, they are embedded close (but not identical) to the dark blue nodes, because they are all part of the subgraph. Second, since the green nodes also serve as the nodes on the bridge that are the closest to the subgraphs, the role similarity between the green nodes and other nodes in the bridge should depend on their distance to the subgraph; our model perfectly captures this by having red nodes being closest to the green nodes, followed by teal, yellow, purple, and finally gray nodes. GraphWave, on the other hand, has embedded all the other bridge nodes with almost equal distance to the green nodes. This shortcoming of GraphWave may be attributed to their use of heat wavelet diffusion to \emph{indirectly} model roles.

\begin{table}[h]
\centering
\begin{tabular}{lclccc}
\Xhline{2\arrayrulewidth}
\multicolumn{2}{l}{Shapes}                      & Method    & \thead{Homoge-\\neity} & \thead{Comple-\\teness} & \thead{Silhou-\\ette} \\ \Xhline{2\arrayrulewidth}
\multicolumn{1}{l|}{\multirow{6}{*}{House}}            & \multirow{5}{*}{\includegraphics[width=0.06\textwidth]{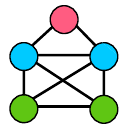}} & node2vec  & 0.005       & 0.005        & 0.330         \\
\multicolumn{1}{l|}{}                                  &                   & RolX      & \textbf{1.000}       & \textbf{1.000}        & \textbf{1.000}          \\
\multicolumn{1}{l|}{}                                  &                   & struc2vec & 0.995       & 0.995        & 0.451         \\
\multicolumn{1}{l|}{}                                  &                   & GraphWave & \textbf{1.000}        & \textbf{1.000}         & \textbf{1.000}        \\
\multicolumn{1}{l|}{}                                  &                   & DRNE & 0.697       & 0.850      & 0.832         \\
\multicolumn{1}{l|}{}                                  &                   & struc2gauss & 0.836       & 0.920      & 0.457       \\
\multicolumn{1}{l|}{}                                  &                   & our method & \textbf{1.000}   &  \textbf{1.000}   &   \textbf{1.000}   \\ \Xhline{2\arrayrulewidth}
\multicolumn{1}{l|}{\multirow{6}{*}{\makecell[c]{House \\ perturbed}}}  & \multirow{5}{*}{\includegraphics[width=0.06\textwidth]{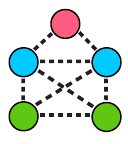}} & node2vec  & 0.030       & 0.032        & 0.276          \\
\multicolumn{1}{l|}{}                                  &                   & RolX      & \textbf{ 0.570}       & 0.588        & 0.346             \\
\multicolumn{1}{l|}{}                                  &                   & struc2vec & 0.206       & 0.235        & 0.180          \\
\multicolumn{1}{l|}{}                                  &                   & GraphWave &  0.547       & 0.566        & 0.374     \\
\multicolumn{1}{l|}{}                                  &                   & DRNE & 0.493      & 0.564      & 0.880       \\
\multicolumn{1}{l|}{}                                  &                   & struc2gauss & 0.162       & 0.323      & 0.005       \\
\multicolumn{1}{l|}{}                                  &                   &our method  &  0.525 &  \textbf{0.603} &\textbf{0.902}  \\ \Xhline{2\arrayrulewidth}
\multicolumn{1}{l|}{\multirow{6}{*}{Varied}}           & \multirow{5}{*}{\includegraphics[scale=0.8]{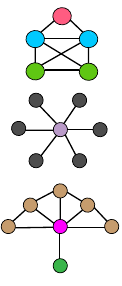}} & node2vec  & 0.244       & 0.216        & 0.400          \\
\multicolumn{1}{l|}{}                                  &                   & RolX      & 0.841       & 0.862        & 0.736        \\
\multicolumn{1}{l|}{}                                  &                   & struc2vec & 0.629       & 0.578        & 0.240          \\
\multicolumn{1}{l|}{}                                  &                   & GraphWave & 0.828       & 0.852        & 0.816     \\
\multicolumn{1}{l|}{}                                  &                   & DRNE & 0.630     & 0.904      & 0.737       \\
\multicolumn{1}{l|}{}                                  &                   & struc2gauss & 0.210       & 0.616      & 0.078       \\
\multicolumn{1}{l|}{}                                  &                   &our method  & \textbf{0.888} &  \textbf{0.938}  & \textbf{0.991} \\ \Xhline{2\arrayrulewidth}
\multicolumn{1}{l|}{\multirow{6}{*}{\makecell[c]{Varied \\ perturbed}}} & \multirow{5}{*}{\includegraphics[scale=0.8]{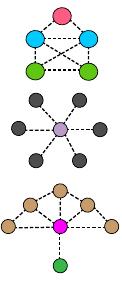}} & node2vec  & 0.303       & 0.265        & 0.360          \\
\multicolumn{1}{l|}{}                                  &                   & RolX      & 0.638       & 0.627        & 0.418    \\
\multicolumn{1}{l|}{}                                  &                   & struc2vec & 0.457       & 0.433        & 0.289          \\
\multicolumn{1}{l|}{}                                  &                   & GraphWave & \textbf{0.697}       & 0.680        & 0.516   \\
\multicolumn{1}{l|}{}                                  &                   & DRNE & 0.488     & 0.651      & 0.728       \\
\multicolumn{1}{l|}{}                                  &                   & struc2gauss & 0.116       & 0.458      & 0.039       \\
\multicolumn{1}{l|}{}                                  &                   &our method  & 0.506   & \textbf{0.713} & \textbf{0.920}  \\ \Xhline{2\arrayrulewidth}
\end{tabular}
\caption{Node clustering results averaged over 25 runs. The best result for each metric is shown in bold. The results of  node2vec, struc2vec, RolX, GraphWave, and the figures in the table are taken from Donnat et al.\cite{graphwave}.  }
\label{cluster}
\vspace{-15pt}

\end{table}

\subsection{Node Clustering}
The next set of experiments involve quantitative evaluations of the generated embeddings through node clustering. For these experiments, we use the same synthetic graphs and metrics used by GraphWave \cite{graphwave}. For evaluation, we use agglomerative clustering with single linkage to cluster embeddings and report the average homogeneity, completeness \cite{rosenberg2007v}, and the silhouette score \cite{rousseeuw1987silhouettes} after 25 runs (the random seeds to generate the cycles are set to 0 at the beginning of the first run). The results are shown in Table \ref{cluster}. For the baselines node2vec, struc2vec, RolX, and GraphWave, we report the results provided by Donnat et al. (\citeyear{graphwave}). Our model, along with RolX and GraphWave, achieves perfect scores for the house setting. In all other settings, our model outperforms all the baselines for all the shapes and metrics with two exceptions (out of 9), where GraphWave and RolX generate better results in the homogeneity metric for the perturbed settings.

\subsection{Node Classification}
\begin{table} 
\centering

\begin{tabular}{llll}
\Xhline{2\arrayrulewidth}
     & Brazilian & American & European  \\
     \Xhline{2\arrayrulewidth}
       our method  & \textbf{0.784} & \textbf{0.657} &\textbf{0.601}  \\
       GraphWave &0.778 & 0.631 & 0.571\\
        DRNE &0.776 &0.578 & 0.533\\
        struc2gauss & 0.314 &  0.351 & 0.310\\
       struc2vec &  0.732 & 0.651 & 0.577   \\
       node2vec &  0.267 & 0.473 & 0.329  \\ 
     
\Xhline{2\arrayrulewidth}
\end{tabular}
  \caption{Average (across 10-fold) micro F1 score of our model vs all the baselines for the node classification task.}
   \label{nodeclassification}
   \vspace{-21pt} 
\end{table}

The final set of experiments involve node classification on three real-world datasets provided by Ribeiro et al. \cite{struc2vec}: Brazilian, American, and European air-traffic networks. The nodes correspond to airports and are labeled with one of four possible labels, based on their activity. We use all the baselines and our method to extract embeddings from each network and then run 10-fold cross-validation using a support vector machine implemented using scikit-learn \cite{scikit-learn}. Table \ref{nodeclassification} shows the node classification results. Our method outperforms all the baselines across all three datasets. This is further quantitative evidence of our method's superiority in embedding node structural roles.

\section{Conclusion}
In this paper, we introduced a novel and flexible structural role embedding framework using stress majorization, which can directly and precisely capture the role structural identities and similarities of nodes in networks. We also provided a strictly mathematical proof that nodes with the same roles overlap perfectly in the embedding space when embedded using our framework.
We validated our method through qualitative and quantitative evaluations on synthetic and real-world datasets, showing that our method outperforms other well-known related methods in learning node role representations, across all tasks. The code and data for this paper will be made available upon request. 
\bibliographystyle{ACM-Reference-Format}
\balance
\bibliography{sample-base}

\end{document}